Invited Paper

# Review of germanium-silicon single-photon avalanche diodes

Neil Na*[a], Erik Chen[a], Gerald S. Buller[b], Robert H. Hadfield[c], and Richard A. Soref[d]

[a]Artilux Inc., Zubei City, Hsinchu County 302082, Taiwan
[b]Institute of Photonics and Quantum Sciences, Heriot-Watt University, Edinburgh EH14 4AS, UK
[c]James Watt School of Engineering, University of Glasgow, Glasgow, G12 8QQ UK
[d]Department of Engineering, University of Massachusetts, Boston, Massachusetts 02125, USA

## ABSTRACT

While it took about a decade for a germanium (Ge) thin film grown on a silicon (Si) substrate to be successfully applied as a detector material for high-speed optical fiber communication application, it took about another decade to further expand its usage as a sensor material for active optical sensing and imaging applications. In this paper, we shall review the progress of a shortwave infrared (SWIR) single-photon detection (SPD) with germanium-silicon (GeSi) single-photon avalanche diode (SPAD), ranging from the first demonstration at cryogenic temperature (Z. Lu et al., 2011) to the recent demonstration at room temperature (N. Na et al, 2024). Potential new applications will also be discussed.

**Keywords:** Ge-on-Si, GeSi, single-photon avalanche diode, single-photon detection, Si photonics, SWIR

## 1. INTRODUCTION

Ge is one of the oldest materials studied in the history of semiconductor research and development. At the electronics side, the first solid-state transistor was demonstrated by Bardeen and Brattain at Bell Labs in 1947 using two closely spaced gold contacts on a Ge crystal, where their work was later awarded with Nobel prize for the invention of the transistor. However, due to factors such as the facts that Ge surfaces are harder to passivate and Ge wafers are more difficult to grow in large diameters, complementary metal-oxide-semiconductor (CMOS) transistors using Si eventually dominated and became a commercial success. Since then, Ge was predominately used as SiGe, an alloy containing a small portion of Ge in a Si matrix, for specialized applications such as bipolar CMOS (BiCMOS) heterojunction bipolar transistor (HBT), or source/drain strain engineering in CMOS transistors.

At the photonics side, Ge has long been used as a transparent material for making lenses, waveguides, and other optics at mid infrared (MIR) wavelengths. At near infrared (NIR) and SWIR wavelengths, there were attempts to fabricate photodiodes (PDs) over Ge wafers, but they never became a commercial success due to similar difficulties in passivating Ge surfaces and growing Ge wafers in large diameters. The obstacle of working with the Ge wafer was solved in the past several decades because of the development of Ge-on-Si technology, in which a high-quality Ge PD can be fabricated by directly growing Ge on the Si wafer [1]. Such a breakthrough was accompanied by the rise of Si photonics (SiPh), in which photonic components, such as Ge waveguide PD or Ge normal-incidence PD, can be fabricated by the CMOS process and at the same time integrated with CMOS electronics: In the year 2010, Intel [2,3] demonstrated a 4-channel coarse-wavelength-division-multiplexing (CWDM) SiPh transceiver (including Ge waveguide PD) using the Si-on-insulator (SOI) substrate, with CMOS electronics wired-bonded to the photonic integrated circuit (PIC); in the year 2011, Luxtera [4] demonstrated essential SiPh components (including Ge waveguide PD) using the SOI substrate, with CMOS electronics monolithically integrated with the PIC; in the year 2014, Samsung [5] demonstrated essential SiPh components (including Ge waveguide PD) using the bulk-Si substrate, with CMOS electronics wired-bonded to the PIC; in year 2018, Artilux [6,7] demonstrated a 240×180 resolution indirect time-of-flight (TOF) CMOS image sensor (including Ge normal-incidence PD) using the Si substrate, with CMOS electronics wafer-level hybrid-bonded to the pixel/micro-lens layer. In all of the examples above, PDs fabricated by Ge-on-Si technology are utilized.

*neil@artiluxtech.com

Note that the optical and electrical properties of Ge-on-Si are different from bulk Ge, which is illustrated in Figure 1: due to the thermal expansion mismatch between Ge and Si, biaxial tensile strain is induced and decreases/decreases/increases the Γ electron/heavy-hole/light-hole energy, respectively, extending the absorption wavelength to ~ 1.6 μm. Moreover, defects arising from the 4.2% lattice mismatch between Ge and Si generate larger dark current compared to its III-V counterpart, e.g., growing indium-gallium-arsenide (InGaAs) on an indium-phosphide (InP) wafer. As a result, the usage of Ge-on-Si has long been limited to high-speed optical communication applications [2,3,4,5], because in this case the system noise is dominated by transimpedance amplifier (TIA) circuit noise rather than detector dark current. On the other hand, applying Ge-on-Si to optical sensing and imaging applications [6,7,8] raises tremendous technical difficulties, because in these applications the system noise is dominated by detector dark current rather than by analog frontend (AFE) circuit electronic noise.

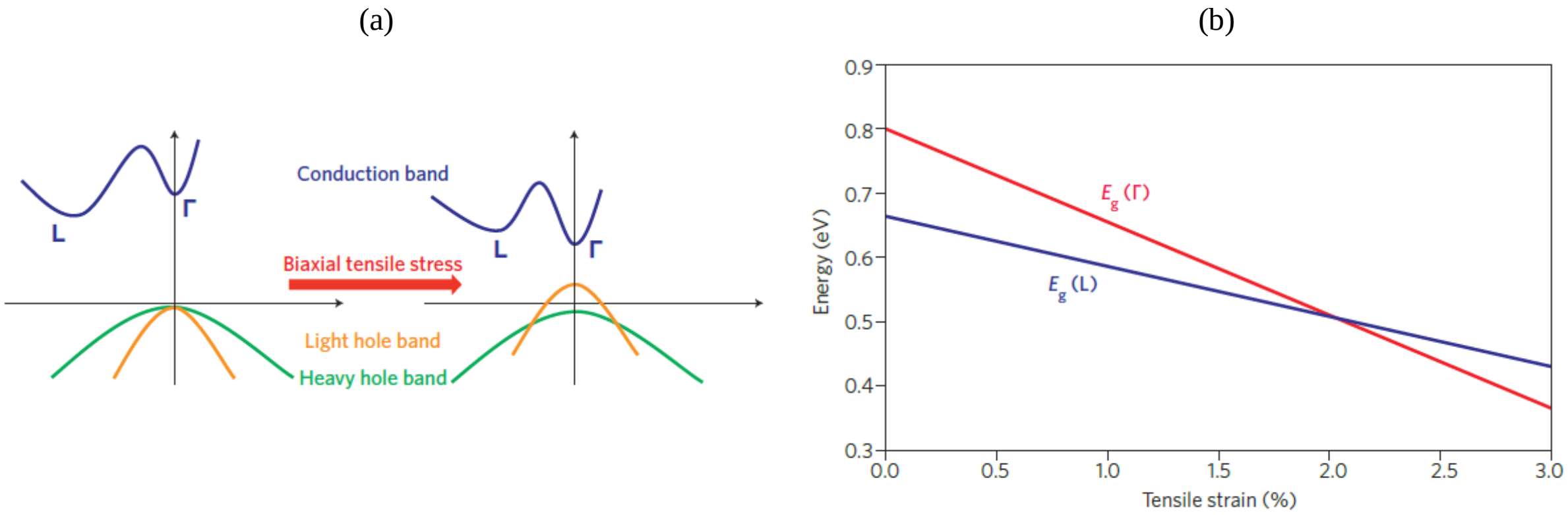

Figure 1. The effect of tensile strain on the band structure of Ge. (a) Schematic of how the band diagram changes as biaxial tensile strain is applied. (b) Plot of the bandgap energies $E_g(\Gamma)$ and $E_g(L)$ as a function of tensile strain. Figures reproduced with permission from Ref. [1] NPG.

Despite the issue of detector dark current, researchers have been working on advancing the techniques of material preparation, such as blanket/selective epitaxy [9], defect-necking epitaxy [10], liquid-phase epitaxy [11], and post-epitaxy chemical-mechanical-polishing (CMP) [12], as well as expanding the types of device demonstration, such as the homojunction Ge PD [13], the heterojunction GeSi avalanche photodiode (APD) [14] (in the following, APD represents the PD operated in the linear mode below the breakdown voltage), and heterojunction GeSi SPAD [15,16,17] (in the following, SPAD represents the PD operated in the Geiger mode above the breakdown voltage). In this paper, instead of presenting a comprehensive review, we highlight the key milestones achieved in the Ge-on-Si technology, from material preparation to device demonstration (focusing on normal-incidence geometry rather than waveguide geometry), leading to the recent demonstration of the GeSi SPAD operating at room temperature [17], and possible future applications beyond communication, sensing, and imaging, such as photonic quantum computing (PQC) [18] at room temperature [19].

## 2. MATERIAL PREPARATION

**Blanket/selective epitaxy**

In the year 1999, MIT [9] demonstrated selective epitaxy of Ge on Si along with post-Ge-growth cyclic anneal to significantly reduce the threading dislocations arising from the lattice mismatch between Ge and Si. In this selective growth experiment, due to the large seed window ranging from 10 to 100 μm, threading dislocations form between the Ge surface and the Ge-Si interface, similar to what is typically observed in a blanket epitaxy experiment. Ultrahigh-vacuum chemical-vapor-deposition (UHVCVD) is used to perform a two-step epitaxy, one at 350 ˚C and the other at 600 ˚C, in order to grow 1 μm thick Ge-on-Si samples. When the cyclic anneal is between 780 ˚C and 900 ˚C, low threading dislocation density ~ $2\times10^6$ $cm^{-2}$ can be observed. Fig. 2(a) shows an etch-pit density ~ $2.1\times10^6$ $cm^{-2}$, which matches very well the measured $2.3\times10^6$ $cm^{-2}$ by plane-view transmission electron microscopy (TEM). In Fig. 2(b), the cross-sectional scanning electron microscopy (SEM) images before and after the cyclic anneal are shown. Such an approach has been widely used in fabricating high-speed homojunction and heterojunction Ge-on-Si detectors ever since.

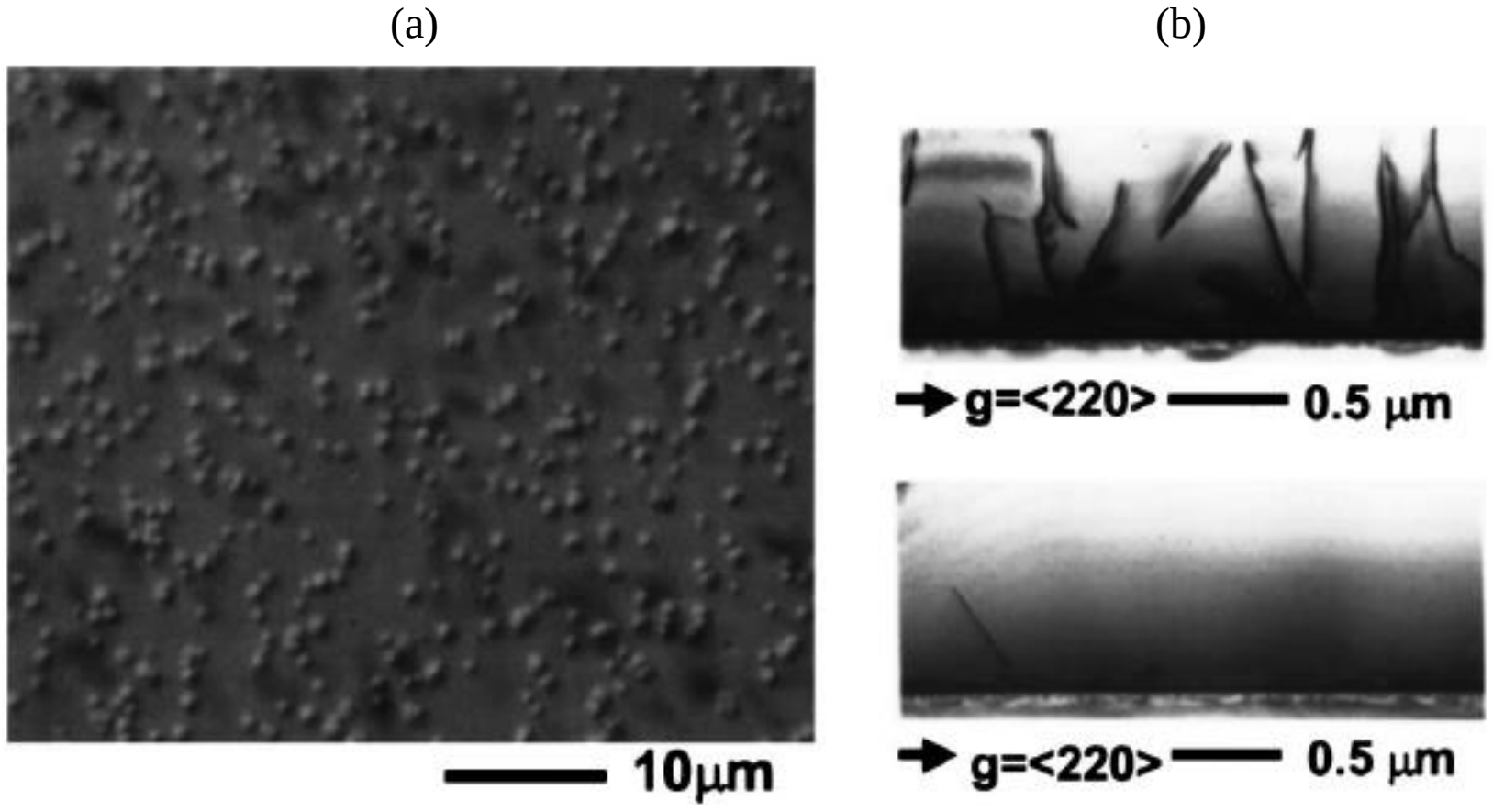

Figure 2. (a) Etch-pit density (EPD) results indicating threading-dislocation density. The average EPD in this figure is $2.1\pm0.5\times10^{7}$ $cm^{-2}$. (b) Cross-sectional TEM image of Ge grown on Si using two-step UHVCVD process (left), and Ge grown on Si using two-step UHVCVD process (right top) followed by 10 cyclic thermal annealing (right bottom). Figures reproduced with permission from Ref. [9] AIP.

### Defect-necking epitaxy

In the year 2000, MIT [10] demonstrated a defect-free top Ge surface by epitaxially necking the threading dislocations with high-aspect-ratio oxide trench. The concept is shown in Fig. 3(a), in which the threading dislocations formed over the (311) plane are terminated by the (010) plane of the oxide trench, resulting in the defect-free top Ge surface. In Fig. 3(b), the defective Ge region in the trench and the epitaxial lateral overgrowth (ELO) region above the trench are shown. Such an approach may produce high-quality Ge-on-Si, and this was applied in fabricating passive SWIR image sensors.

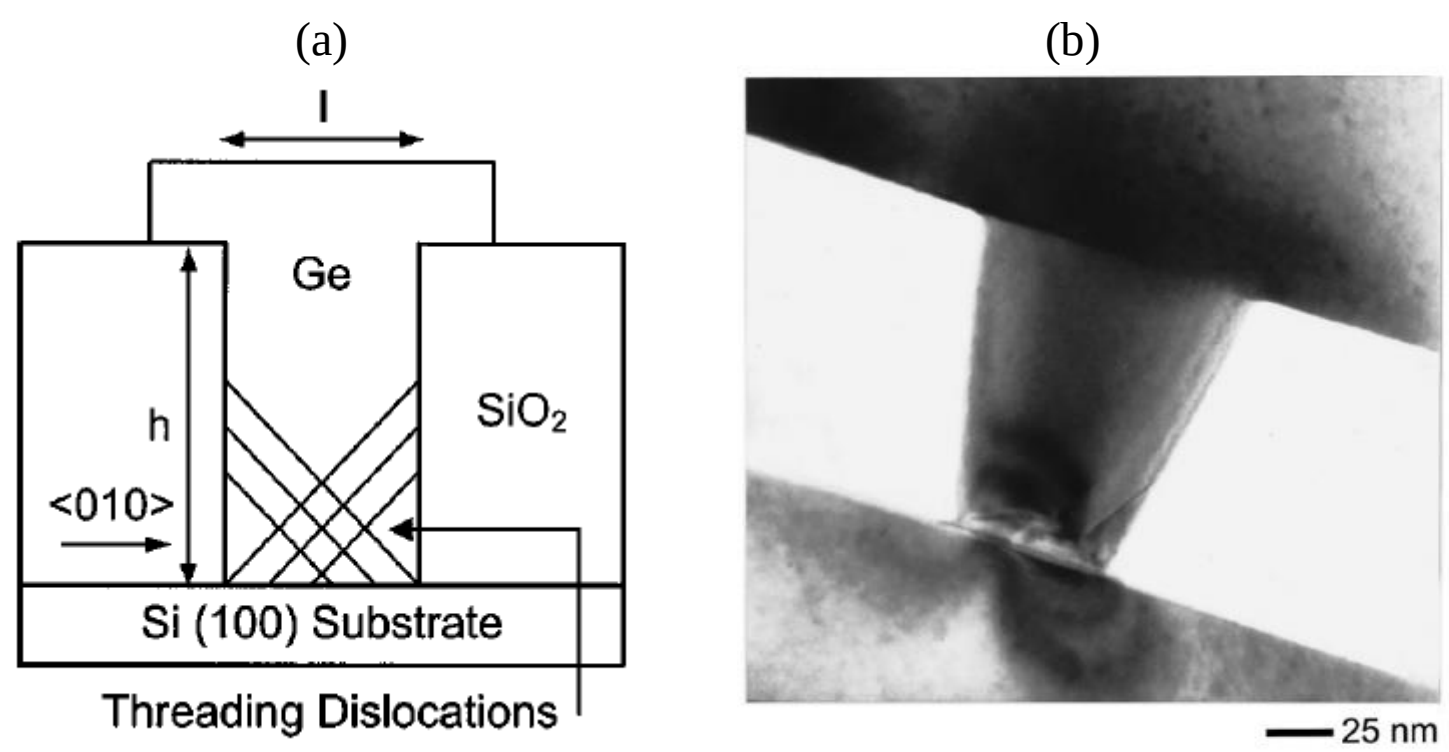

Figure 3. (a) Cross-section diagram demonstrating the principles of epitaxial necking showing zero threading dislocations at the Ge film surface. (b) Cross-section TEM image showing stacking fault blocked by the oxide sidewall and defect free Ge seed and ELO regions. Figures reproduced with permission from Ref. [10] AIP.

**Liquid-phase epitaxy**

In the year 2014, Stanford [11] demonstrated liquid-phase epitaxy (LPE) of Ge on Si. By encapsulating the sputtered Ge film in insulating materials, e.g., nitride and oxide, and with the Si seed region exposed to the sputtered/encapsulated Ge film, the structure serves as a micro-crucible to hold the melted Ge when the wafer is heated up to 940 °C for 2 seconds. Afterwards, the melted Ge is re-crystalized from the Si seed region as the heat is drained from the Si seed region to the Si substrate, and that confines the threading dislocations only in the Si seed region. The schematic plot is shown in Fig. 4(a), and the grown Ge-on-insulator is shown in Fig. 4(b). Such an approach produces high-quality Ge stripes, and was used in fabricating gate-all-around (GAA) Ge-on-insulator transistors.

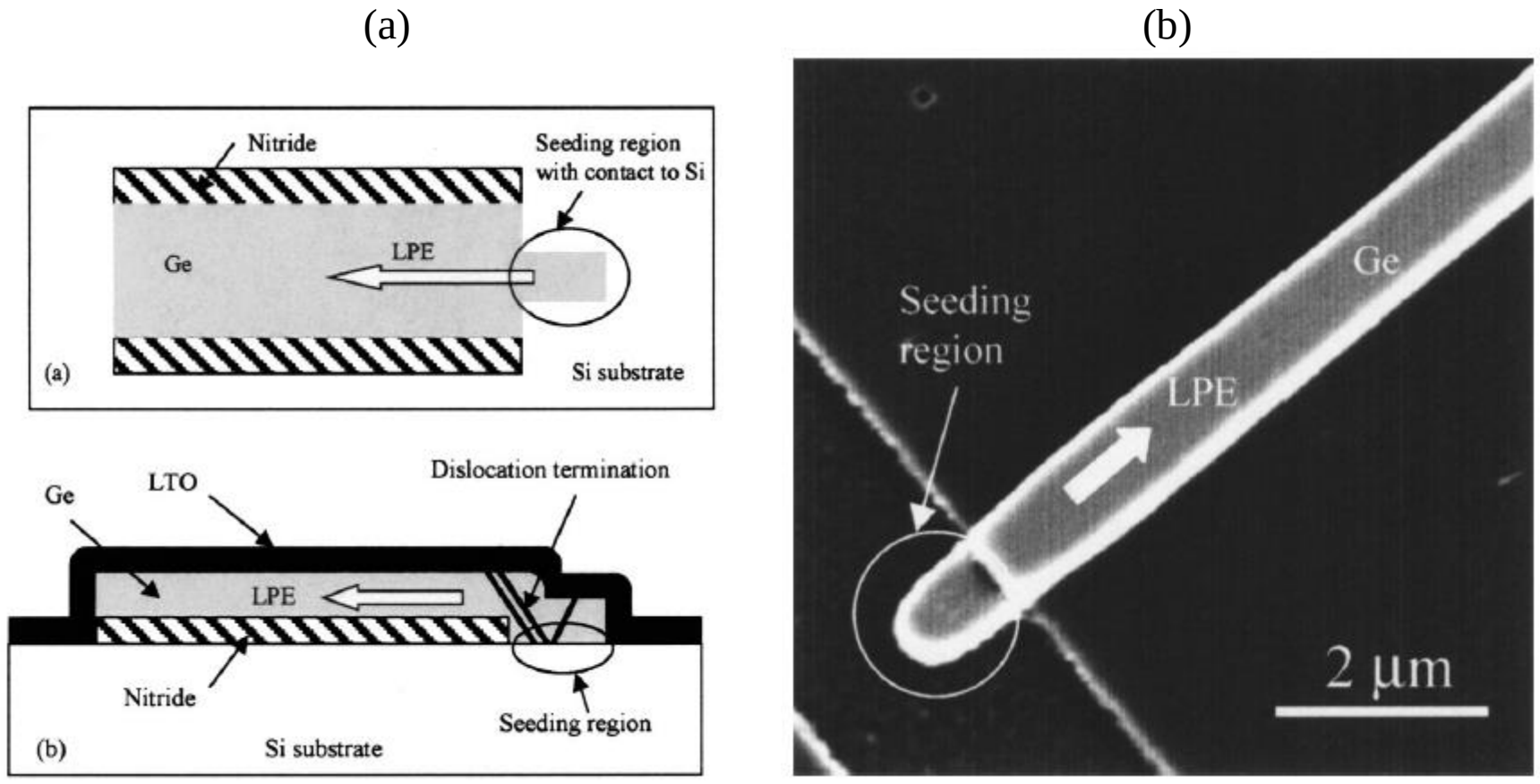

Figure 4. (a) Top view and the cross-sectional schematics of the structure used for Ge LPE growth. The seeding region is expected to be defective due to the lattice mismatch between Ge and Si, but the dislocations grow along the crystallographic planes and terminate quickly. (b) SEM image of the LPE Ge on Si nitride with the seeding window on the Si substrate. Figures reproduced with permission from Ref. [11] AIP.

In all three of the epitaxy growth techniques discussed above, the defective regions can be further removed by various etching techniques so that the device is built upon high-quality Ge. As an example, in Ref. [12], Ge-on-Si is grown by blanket epitaxy with post Ge thermal anneal. After flipping and attaching the Ge-on-Si wafer over a carrier wafer, the threading dislocations are removed by CMP of the backside of the Ge-on-Si wafer, all the way down to the Ge-Si interfacial region.

## 3. DEVICE DEMONSTRATION

**Normal-incidence Ge PD**

Around the year 2010, there are numerous works done on Ge PD in the literature, but very few of them focus on the receiver and/or the system. In the year 2009, Intel [13] first demonstrated a high-speed Ge PD wire-bonded to a TIA in order to compare the eye diagram and the sensitivity with those of a high-speed GaAs PD wire-bonded to a TIA. In that work, the normal-incidence Ge PD consists of a vertical n-i-p junction, in which the n-doping region faces the illumination side. See Fig. 5(a) for the device schematic plot. The n-doping region covers the surface amorphous Si (a-Si) layer and a portion of the surface Ge, while the p-doping region covers the interfacial Si substrate and a portion of the interfacial Ge. This way, the high field region is confined in Ge and the device can be treated as a homojunction Ge PD. The thickness of the Ge film is 1.2 µm and the diameter of the Ge PD is 50 µm. At room temperature, dark current density is measured to be around 6 mA/cm$^{-2}$ at 1-V of reverse bias, responsivity is measured to be nearly 0.6 A/W at 850 nm wavelength and 1-V of reverse bias, and bandwidth up to 9 GHz is measured at 850 nm wavelength and 2-V of reverse bias. While the dark current of the Ge PD is about three orders of magnitude greater than its commercial III-V counterpart, such a large dark current does not influence the receiver performance, i.e., the eye diagram and the sensitivity measured with the receivers using Ge PD or the GaAs PD are identical, as shown in Fig. 5(b). As explained previously, this is due to the high TIA circuit noise associated with high-speed optical communication application.

(a) (b)

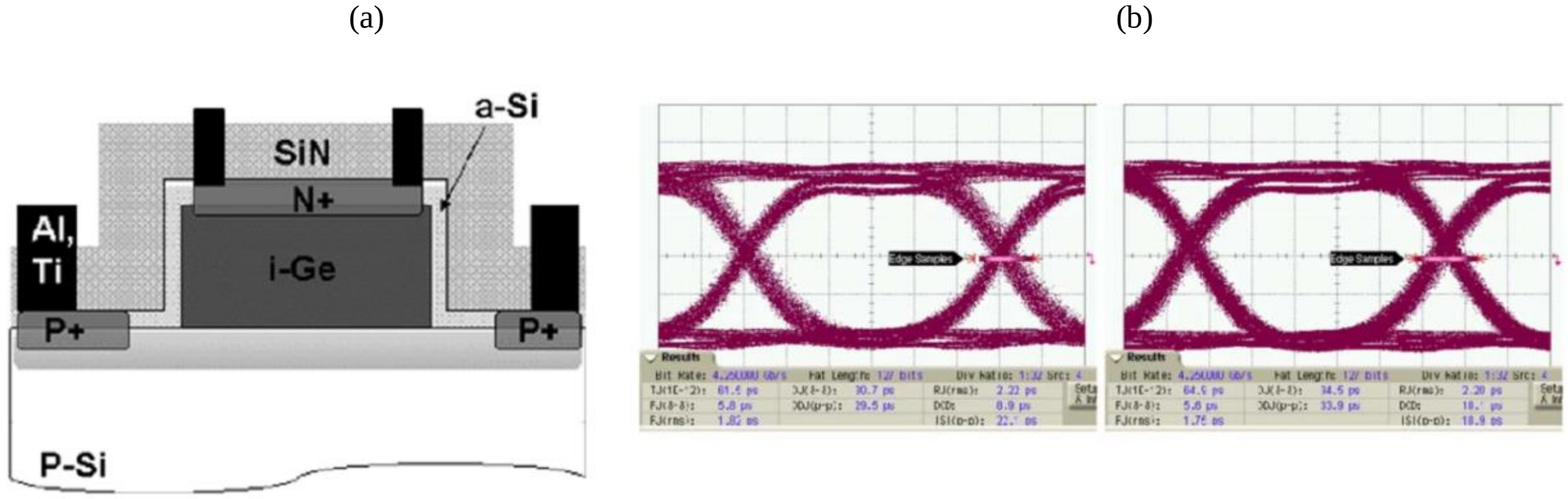

Figure 5. (a) The structure of the Ge PD. (b) Eye diagrams at 4.25 Gb/s and -12 dBm for receivers containing a GaAs PD (left) and Ge PD (right). The jitters of the two receivers are similar and well within the specification for commercial receivers. Figures reproduced with permission from Ref. [13] IEEE.

### Normal-incidence GeSi APD

In the same year 2009, Intel [14] first demonstrated a Ge/Si (abbreviated as GeSi in the following) APD with a $k_{eff}$ of 0.09 and a gain-bandwidth-product (GBP) of 340 GHz, which is the highest ever demonstrated for operation at 1310 nm wavelength. See Fig. 6(a) for the device schematic plot. The thickness of the Ge film is 1.1 µm and the diameter of the Ge APD is 30 µm. The device is based on separate-absorption-charge-multiplication (SACM) design, in which Ge is used for optical absorption and Si is used for electrical multiplication. This way, the high field region exists in both the Ge and Si regions and drives the transports of electrons and holes across the interface between Ge and Si, and so the device can be treated as a heterojunction Ge APD. The optical receivers built with such a GeSi APD wire-bonded to a TIA show a sensitivity of -28 dBm at 10 Gbps, as shown in Fig. 6(b), which is the first reported GeSi APD receiver that has a sensitivity performance comparable to that of the commercial III-V counterpart. Since then, the SACM design becomes the main structure of focus for GeSi SPAD research and development, and will be discussed in the next paragraphs.

(a) (b)

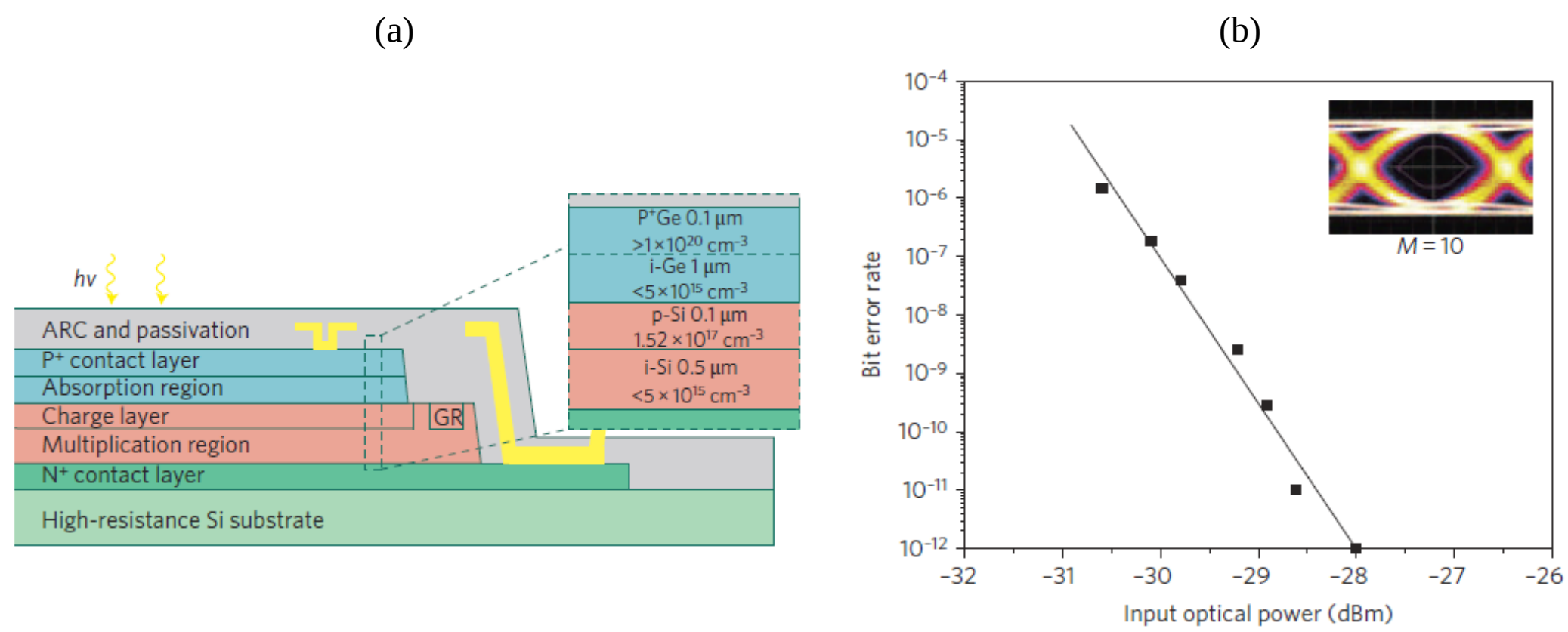

Figure 6. (a) The structure of the GeSi APD. (b) Back-to-back receiver sensitivity and eye diagram measurements for a 30-µm-diameter GeSi APD receiver measured at 10 Gbps. The BER is measured at the optimal bias voltage with 10 Gbps NRZ PRBS optical pulses at 1304 nm. The extracted receiver sensitivity is -28 dBm at a gain of 19, BER of $1\times10^{-12}$ and ER of 12 dB. The inset is an eye diagram at a gain of 10 at -20 dBm input optical power. Figures reproduced with permission from Ref. [14] NPG.

### Normal-incidence GeSi SPAD

The first demonstration of SPD using Ge-on-Si is reported in the year 2011 by the University of Virginia [15]. The same type of SACM samples in Ref. [14] are applied for this experiment, but operated in Geiger-mode instead of linear mode. See Fig. 7(a) for the device schematic plot. Using gated-mode quenching at 200 K, the dark count rate (DCR) is measured to be ~ 400 MHz, and the single photon detection efficiency (SPDE) is measured to be 14 %, as shown in Fig. 7(b). After-pulsing is characterized by DCR versus frequency at 200 K, and, in the frequency range from 1 kHz to 1 MHz, no significant changes are observed. The jitter at 1 V excess bias at 200 K is measured to be 195 ps with 1 photon per pulse. While this is the first demonstration of the GeSi SPAD, the large dark current and hence the large DCR saturate the gated-mode quenching, and so SPDE cannot be faithfully measured unless the system is cooled down to 200 K or lower to alleviate the DCR saturation.

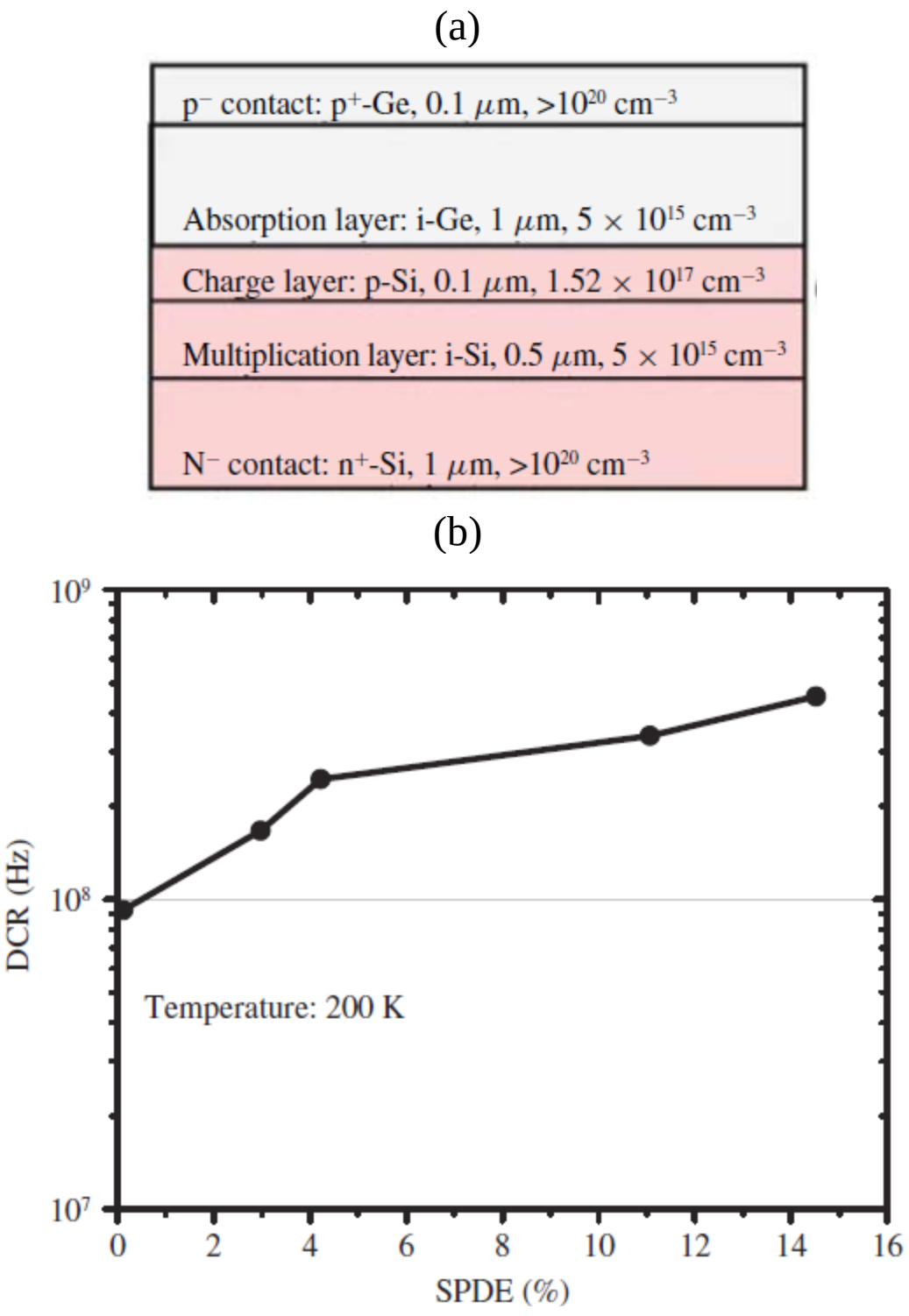

Figure 7. (a) The structure of the GeSi SPAD in Ref. [15]. (b) SPDE versus DCR at 200 K. The excess biases range from 0.5 to 1.02 V. Figures reproduced with permission from Ref. [15] IEEE.

A higher-performance GeSi SPAD with its charge layer and therefore electric field confined to its center with a planar geometry is first proposed and demonstrated by the Heriot-Watt University in year 2019 [16]. See Fig. 8(a) for the device schematic plot. The thickness of the Ge film is 1 µm and the diameter of the GeSi SPAD is 100 µm. The GeSi SPADs features a SPDE of 38 % and a DCR of 2 MHz at T = 125 K at 1310 nm wavelength, as shown in Fig. 8(b). A significant improvement of the noise-equivalent power (NEP) down to $1.9\times10^{-16}$ W/sqrt(Hz) at T = 78 K is achieved, which is a 50-fold improvement over the previously reported GeSi SPADs. After-pulsing performance is analyzed using the time-correlated carrier counting method, observing at least a 50 % reduction in dead-time compared to commercial InGaAs/InP SPADs under the same operating conditions, leading to much higher maximum count rates. This experiment opens the possibility of applying the CMOS-compatible GeSi SPAD for single-photon three-dimensional (3D) sensing and imaging at eye-safe SWIR wavelengths, dose monitoring for laser medicine, as well as for a range of quantum technologies at telecommunication 1310 nm and 1550 nm wavelengths.

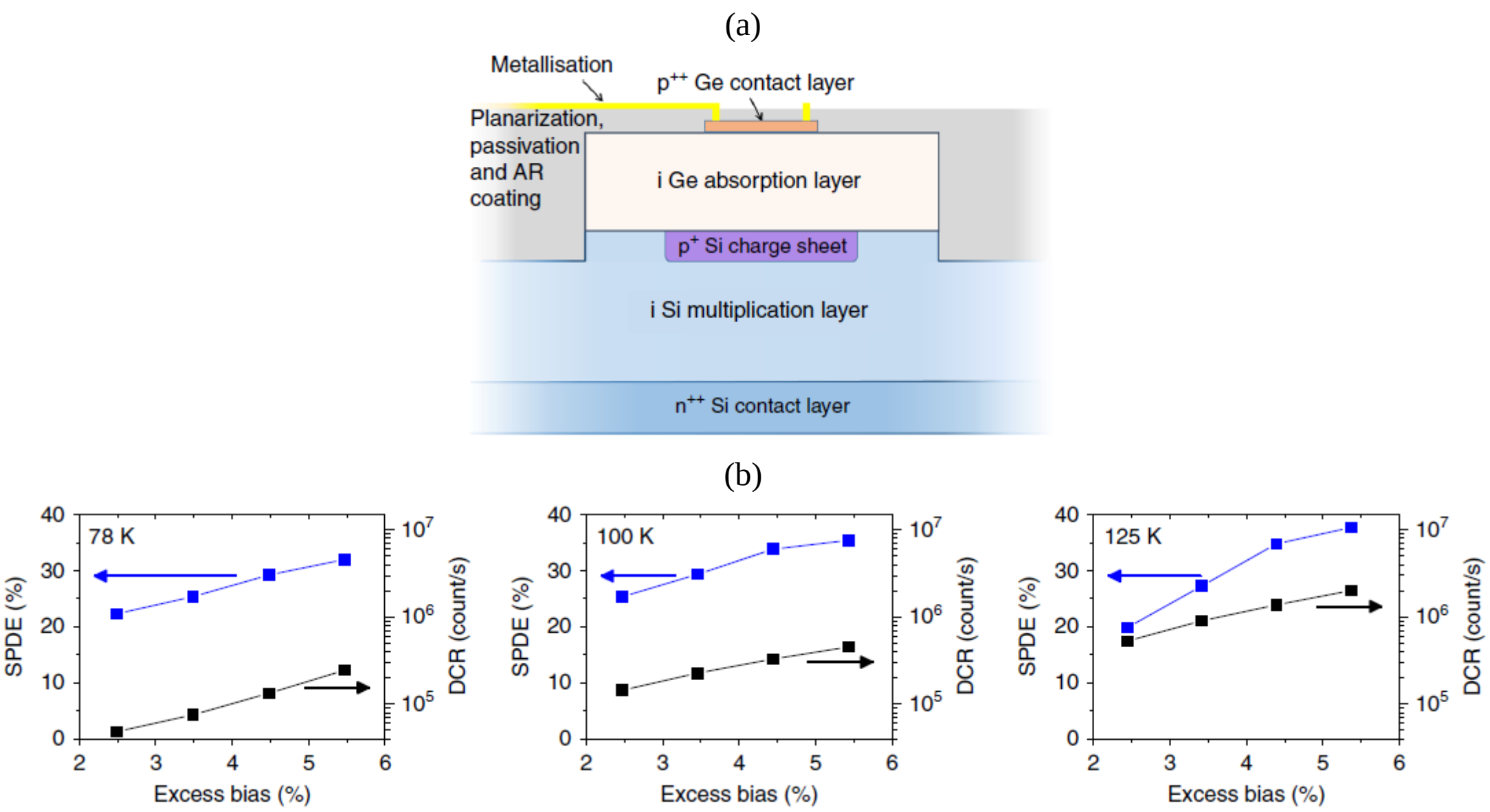

Figure 8. (a) The structure of the GeSi SPAD in Ref. [16]. (b) SPDE and DCR at 78 K (left), 100 K (middle), and 125 K (right). Figures reproduced with permission from Ref. [16] NPG.

In the year 2024, a high-performance GeSi SPAD operating at room temperature was first reported by Artilux [17], featuring a NEP improvement over the previous GeSi SPADs by 2 to 3.5 orders of magnitude. See Fig. 9(a) for the device schematic plot. The device utilizing doping designs to screen the low-quality material regions from the electric field so that dark current and hence DCR can be minimized. Key parameters such as DCR, single-photon detection probability (SPDP) at 1310 nm, timing jitter, after-pulsing characteristic time and after-pulsing probability are, respectively, measured as 19 kHz/μm$^2$, 12%, 188 ps, ~ 90 ns and < 1%, with a low breakdown voltage of 10.26 V and a small excess bias of 0.75 V. The thickness of the Ge film is 450 nm and the diameter of the GeSi SPAD is 10 μm. For the first time, 3D point-cloud images are captured with direct TOF technique using the GeSi SPAD operating at room temperature. This work paves the way towards using single-photon-sensitive SWIR sensors, imagers and PICs in everyday life. Future directions include scaling the device area for developing a sensing and imaging pixel at a pitch smaller than 10 μm, and heterogenous integration with CMOS circuits through wafer-level hybrid-bonding.

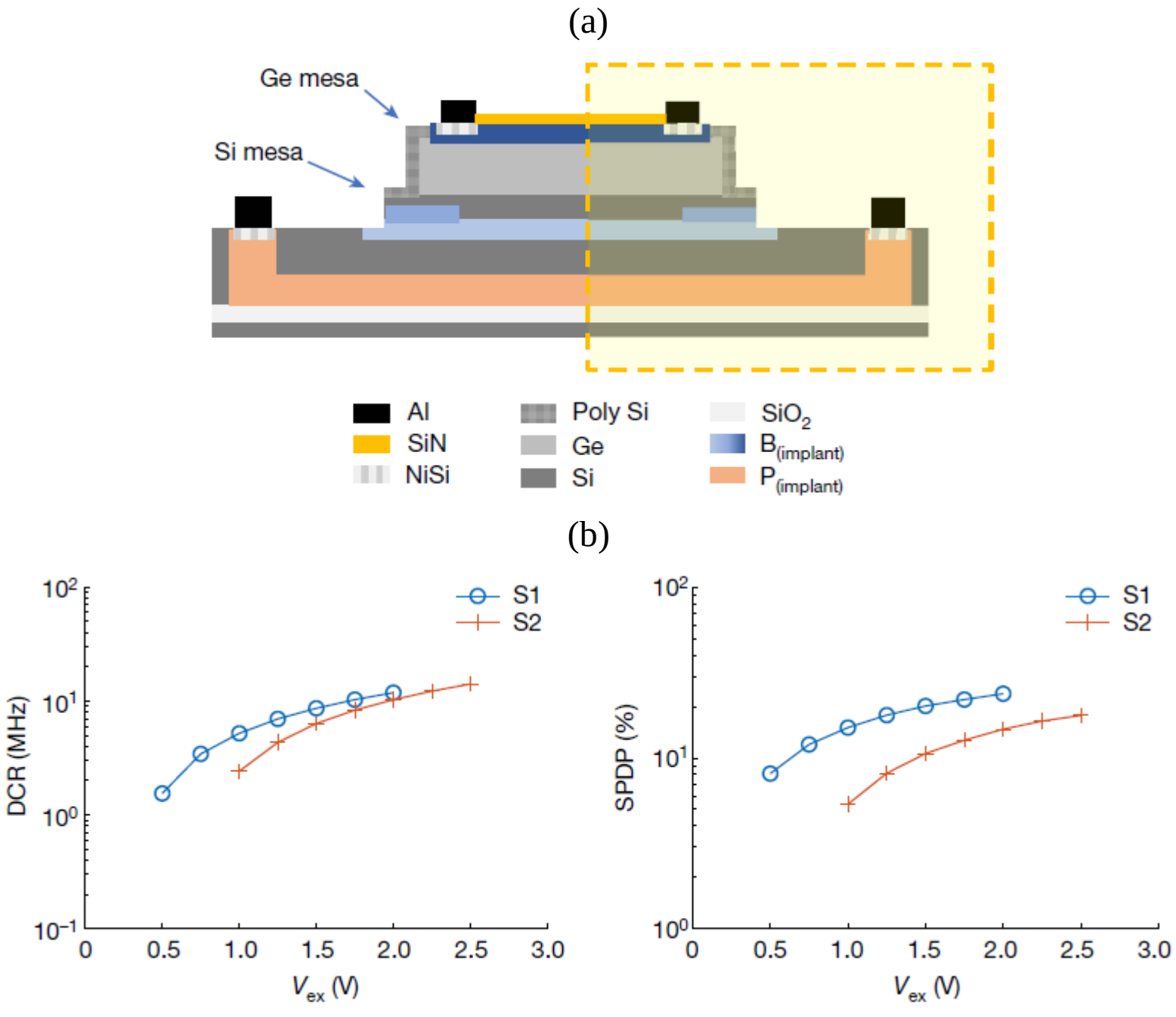

Figure 9. (a) The structure of the GeSi SPAD in Ref. [17]. (b) DCR (left) and SPDP (right) at 300 K. Figures reproduced with permission from Ref. [17] NPG.

## 4. APPLICATIONS BEYOND OPTICAL COMMUNICATION, SENSING, AND IMAGING

So far, GeSi PD/APD/SPAD are used in applications limited to optical communication, sensing, and imaging applications. The demonstration of the room-temperature operated GeSi SPADs indicates that a new application, photonic quantum computing (PQC) at room temperature, may be realized. See Fig. 10 for the system schematic plot. This is due to the fact today that most, if not all, PQC relies upon superconducting nanowire single-photon detectors (SNSPDs) typically based on niobium nitride (NbN) operated at a temperature < 4 K [18]. However, as shown in Ref. [19], for the cases of gate-based PQC and number-based PQC, the performance of the waveguide GeSi SPAD at room temperature can be competitive against that of the first-generation NbN SNSPDs at cryogenic temperature [20], in a series of metrics for PQC with a reasonable time-gating window. Moreover, these GeSi SPADs become photon-number-resolving avalanche diodes (PNRADs) by deploying a spatially-multiplexed M-fold-waveguide array of M GeSi SPADs. These findings may pave a new path toward cryogenics-free PQC that will significantly increase the testing throughputs and reduce the design iteration cycles, which are essential for PQC to continuously evolve to full fruition.

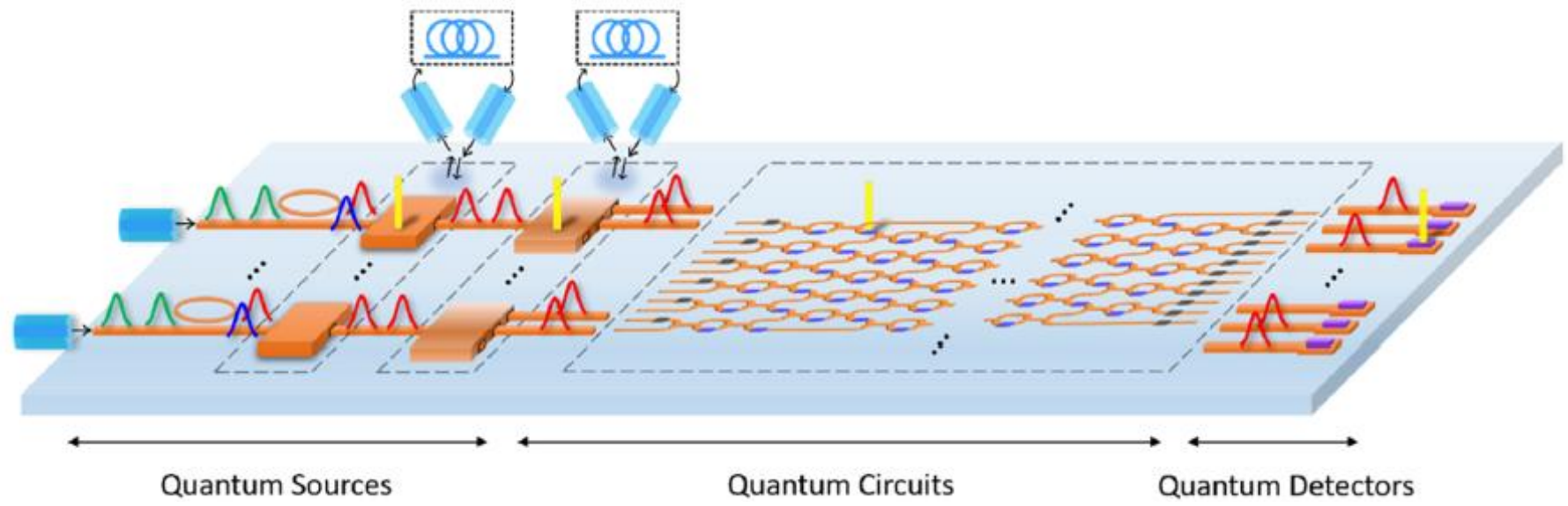

Figure 10. (a) Schematic plot of a room-temperature PQC paradigm with integrated SiPh using the path degree of freedom of single photons: single photons are generated through SFWM (green pulses converted to blue and red pulses) in SOI rings (orange circles), followed by active temporal multiplexers (orange boxes that block the blue pulses), and active spatial multiplexers (orange boxes that convert serial pulses to parallel pulses) (quantum sources), manipulated by a field-programmable interferometer mesh using cascaded Mach-Zehnder interferometers (quantum circuits), and measured by the proposed waveguide GeSi SPADs as single-photon and/or number-photon detectors (quantum detectors). The off-chip fiber couplings are either for the pump lasers or the optical delay lines. Figures reproduced with permission from Ref. [19] AIP.

## 5. CONCLUSION

Through the research and development of Ge-on-Si technology in the past two decades, Ge-based detectors have now been widely applied in optical communication applications, and just start to penetrate optical sensing, imaging, and computing applications. This review summarizes the key milestones of Ge-on-Si technology, starting from material preparation to device demonstration, and ultimately leading to the GeSi SPAD operating at room temperature. Future direction may include extending the detection wavelength of GeSi SPAD beyond 1.6 μm using GeSn-Si SPAD [21] toward room temperature operation.

## ACKNOWLEDGEMENTS

E.C. would like to thank Prof. Edoardo Charbon for a fruitful discussion. R.S. is supported by the U.S. Air Force Office of Scientific Research under Grant No. FA9550-19-1-0341.